\documentclass{article}
\usepackage{amsmath}
\usepackage[english]{babel}
\usepackage[utf8]{inputenc}
\usepackage{graphicx}
\usepackage{soul}
\usepackage[bottom=2.5cm,left=3cm,right=3cm,top=3cm]{geometry}

\begin{document}
\begin{center}
\section*{\huge Vertical Bicycle Model of Vehicle with Dry  Friction  }
\textit{G.M. Rozenblat$^{1}$}\\
$^{1}~$Moscow Automobile and Road Construction State Technical University\\
e-mail: gr51@mail.ru

\textit{V.B. Yashin $^{2}$}\\
$^{2}~$Moscow Technical University of  Communications and Informatics\\
e-mail: hekkoki@gmail.com

\end{center}

{\bf Abstract.} {\small The motion of a vertically positioned bicycle is considered when
 a horizontal control force, which maybe both internal and external in relation to the bicycle, 
is applied to its pedal. Tangential forces of dryfriction obeying the Euler -- Coulomb law act at points
 of contact of the wheels with the horizontal supportplane. The constraint at the points of contact  of the wheels 
with the support is assumed to be unilateral.The problem of determining the acceleration 
of the centre of mass of the bicycle and the realized motionsof its wheels (with or without slip, and with or without detachment) 
with dierent values of the design parameters and control force is solved. 
This approach is usefull for control of vehicle  and optimization  of its dynamical parameters.
 Cases of non-uniqueness of motion  the Painleve paradox  are found.}

\section*{Introduction}
The movements of bodies by means of change in the positions of internal (relative to the body) masses occur on exposure to external forces, which generally are friction forces (resistance forces, hydrodynamic forces, gravitational forces). The creation of such effective (optimum) external forces is due to the corresponding synthesis of internal forces on exposure to which admissible movements of internal masses are realized (Ref. $1,$ p. 344). Such problems have been examined (see, for example, Refs $2-7$) for the purposes of creating robotic devices. In some problems, internal forces are used to create a variable shape of a body moving in liquid, [{8--10]. In space dynamics, internal forces can be used for controlled change in the external gravitational force, which can move such a body away from the Earth by as great a distance as desired (the 'graviflight' effect, see Ref. $11,$ Essay 9).

The movement of a bicycle (and any self-propelled vehicle) also belongs to this class of problems of the movement of bodies. Here, the internal (control) force is the force applied to the pedal by the cyclist's foot. External forces, which realize the motion of the bicycle, are the gravitational forces and the forces of reactions at the points of contact of the wheels with the road.

\section{The formulation and solution of the two simplest problems of bicycle motion}
Figure 1 depicts a bicycle moving in the vertical position relative to a stationary right-handed coordinate system Oxyz which is selected in the following way. The $x$ axis coincides with the horizontal support line along which the bicycle is moving and is oriented to the right, and the $y$ axis is directed down the vertical. Thus, the vertical plane coincides with the plane of the bicycle and with the plane $O x y.$ The z axis is perpendicular to the $O x y$ plane and forms with the $x$ and $y$ axes a right-handed coordinate system. Thus, the $z$ axis is perpendicular to the plane of the figure and is directed away from the reader. With this choice of coordinate system, all the turns, angular velocities, and angular accelerations of the bodies in the figure that occur in a clockwise direction are considered to be positive. It is clear that these turns, angular velocities, and angular accelerations, when viewed from the positive direction of the $z$ axis, are oriented anticlockwise in this case and are positive, as assumed in theoretical mechanics for three-dimensional space. Similar properties are also achieved for the moments of forces: if a force turns the body clockwise about a selected point, from the point of view of the reader looking at the figure, then the moment of this force relative to the examined point is positive, and in the opposite case it is negative. Below, unless stipulated otherwise, we will assume that the introduced kinematic parameters are positive in the sense indicated above. The true signs of these quantities are obtained as a result of solving the problem.

\begin{figure}
  \centering
  \includegraphics[scale=0.5]{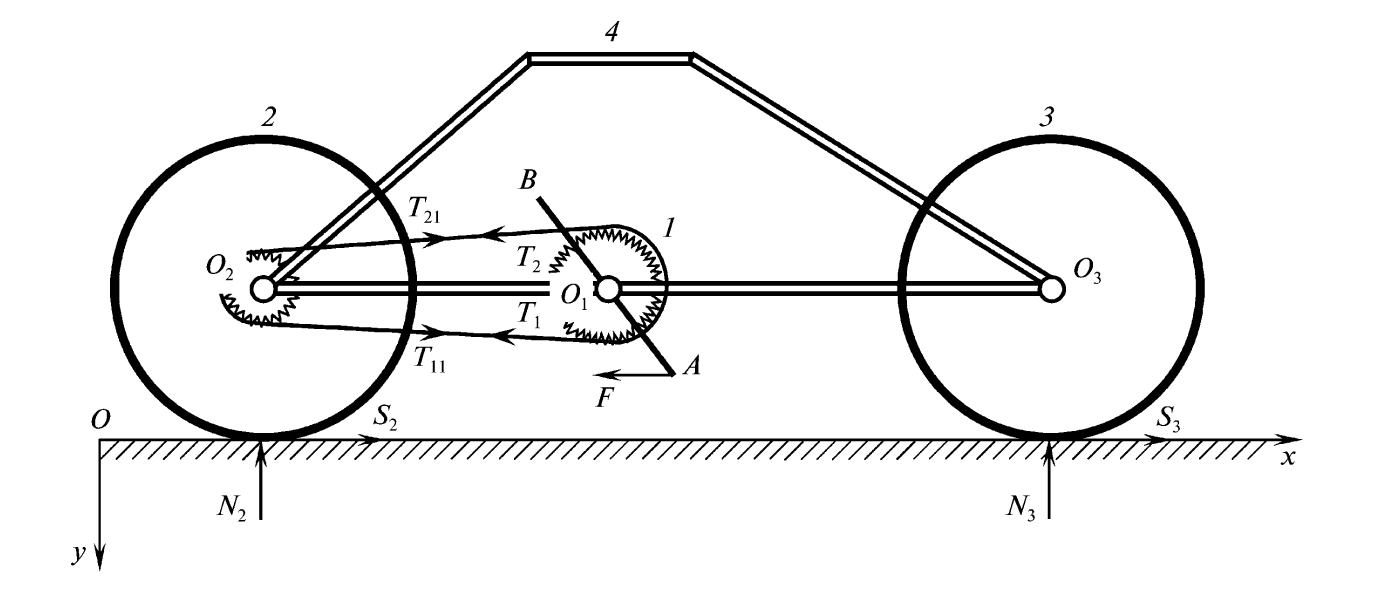}\\
  \caption{}\label{fig1}
\end{figure}

A bicycle consists of four bodies: \emph{1} -- the sprocket with the pedals; \emph{2 and 3} -- the two wheels; \emph{4} -- the frame to which the bodies 1,2, and 3 are hinged at points $O_{1}, O_{2},$ and $O_{3},$ as shown in Fig. $1.$ It is assumed that the cyclist moving the bicyle along the horizontal $x$ axis by applying to the lower pedal $A$ an internal horizontal force $F$ oriented against the direction of the $x$ axis is also rigidly connected to the frame.

We will introduce the following notation: $v$ and $a$ are the velocity and the acceleration of the centre of mass of the bicycle along the $x$ axis, $\omega_{i}$ and $\varepsilon_{i}$ are the angular velocities and accelerations of the sprocket $(i=1),$ the rear wheel $(i=2),$ and the front wheel $(i=3), I_{i}(i=1,2,$
3) is the moment of inertia of the body $i$ relative to the inherent centre of mass, $M=1$ is the total mass of the bicycle (including the cyclist when seated on the bicycle $, r_{1}$ is the radius of the sprocket with the pedal, $r_{2}$ is the radius of the sprocket on the rear wheel, $\mu=r_{2} / r_{1}$ is the gear ratio, $R$ is the external radius of the wheels 2 and $3, N_{j}$ and $S_{j}(j=2,3)$ are the normal reaction and the tangential reaction at the point of contact of the wheel $j$ with the road, $b_{j}(j=2,3)$ is the distance from the point of contact of the wheel $j$ to the centre of mass of the bicycle along the horizontal, $b=\left(b_{2}+b_{3}\right) / 2$ is the half-distance between the points of attachment of the wheels along the horizontal, $h_{0}$ is the elevation of the centre of mass of the bicycle (together with the cyclist when seated on the bicycle) above the road, $F$ is the force applied by the cyclist to the lower pedal of the sprocket 1 and directed along the horizontal against the positive direction of the $x$ axis, $L$ is the corresponding arm (i.e., the distance from the pedal to the horizontal $O_{1} O_{2}$), and $T=T_{1}+T_{2}$ is the difference in the moduli of the forces of tension of the chain that are applied to the sprocket 1.

Let us examine the following problems (below it is assumed that there is no slip of the wheels).

\textbf{Problem 1.} In the standard situation, when the force $F$ applied by the cyclist is internal relative to the indicated system of four bodies, determine the acceleration of the centre of mass of the bicycle (together with the cyclist) along the $x$ axis.

\textbf{Problem 2.} In the non-standard situation, when the force $F$ is external relative to the indicated system of four bodies (the cyclist has dismounted from the bicycle and is applying force $F$ to the lower pedal $A$ against the direction of the $x$ axis or is employing the services of an external aid), determine the acceleration of the bicycle along the $x$ axis.

In particular, in both cases, if before the application of force $F$ the bicycle is at rest, then the side to which it will begin to move - in the direction of force $F$ or in the opposite direction to this force - must be indicated.

\subsection{Solutions of the problems}
We will use the theorem of change in kinetic energy (Ref. 12, p. 69). Assuming that the wheels are rolling without slip, and the drive chain of the bicycle is unstretchable, we will obtain the following kinematic relations:
\begin{equation}\label{eq11}
 \omega_{1}=\mu v / R, \quad \omega_{2}=\omega_{3}=v / R ; \quad \mu=r_{2} / r_{1}
\end{equation}

The kinetic energy of the bicycle, with the use of relations (1.1), is calculated by means of the formula
\begin{equation}\label{eq12}
  T=\frac{1}{2} M_{0} v^{2} ; \quad M_{0}=1+\frac{I_{1} \mu^{2}+I_{2}+I_{3}}{R^{2}}
\end{equation}

To calculate the power of all internal and external forces, we will consider that in Problem 1 a non-zero contribution is made by the power of the internal force $F$, which creates the moment $F L$ applied to the body 1 rotating at angular velocity $\omega_{1}$. In Problem 2, a non-zero contribution is made by the external force $F$ applied to the lower pedal of the bicycle $A$ having the following projection of absolute velocity onto the $x$ axis:
\begin{equation}\label{eq13}
  v_{A x}=v-\omega_{1} L=v(1-\sigma) ; \quad \sigma=\mu L / R
\end{equation}

The remaining (both internal and external) forces make a zero contribution owing to the absence of slip of the wheels, the absence of stretchability of the drive chain, and the ideal hinges at points $O_{1}, O_{2}, O_{3}$. Thus, the power is determined by the expression
$$
W_{1}=F L \omega_{1}=F \sigma v \text { for Problem } 1, \quad W_{2}=-F U_{A x}=-F v(1-\sigma) \text { for Problem } 2
$$

We will describe in greater detail how the formula for power $W_{1}$ is obtained. We will transfer to point $O_{1}$ the force $F$ applied to the pedal $A$. As a result we obtain the force $F$ applied at point $O_{1}$ and a pair of forces with moment $F L$. This system of forces is applied to the body 1 rotating about point $O_{1}$ and moving forwards together with this point. According to Newton's third law, on the body 4 (on whose side this system of forces is applied) at point $O_{1}$ is acted upon by a force $F$ and a pair of forces with moment $F L$. The body 4 does not rotate. and therefore the power of the latter moment is zero, and the two oppositely directed and applied forces at point $O_{1}$ together provide zero power. Thus, a non-zero contribution is made only by the power of the indicated pair of forces applied to the body 1 rotating at angular velocity $\omega_{1}$, and we arrive at the expression given above for power $W_{1}$. Then, using the theorem of change in kinetic energy, we will obtain
\begin{equation}\label{eq14}
  a=\sigma F / M_{0} \text{ for Problem 1}
\end{equation}
\begin{equation}\label{eq15}
  a=-(1-\sigma) F / M_{0} \text{ for Problem 2}
\end{equation}

The parameters $M_{0}$ and $\sigma$ were defined, respectively, by the second equality of system (1.2) and the second equality of system (1.3). From this it follows that in problem 1 the bicycle moves along the $x$ axis, against the force $F$, and in problem 2 with $\sigma \in(0,1)$ (with $\sigma>1$) the bicycle moves against (along) the $x$ axis, in the direction (against the direction) of the force $F$. For normal designs of bicycles, $r_{2}<r_{1}(\mu<1)$, $L<R.$ Then $\sigma \in(0,1),$ and the bicycle moves in the direction of the external force $F$.

\textbf{Remark 1.} Similar problems for a bicycle (three-wheel) on a qualitative level were proposed at the All-Union Physics Olympiads in the $1960 \mathrm{~s}$ (see Ref. 13, problems 98 and 106). As can be seen, the strict and complete solution is fairly complex (for the student), and the sound solution of Problems 1 and 2 requires physical intuition.

\textbf{Remark 2.} The excellent book by Arnold ${ }^{14}$ contains the essay 'Which force drives the bicycle forwards?'. The author applies a backward force to the lower pedal of the bicycle and shows that the bicycle will move forwards. As can be seen, this outcome contradicts the answer to Problem 2 (with adherence to the inequality $0<\sigma<1$, as occurs for real bicycle models). The present author learnt of this discrepancy from V.F. Zhuravlev, and, in part, it prompted the author to write the present paper. It may be obvious (see Fig. 1) that the force $F$ creating a clockwise moment applied to the sprocket will rotate the latter likewise in a clockwise direction. The sprocket, in turn, with the help of the chain, will force the wheels of the bicycle (in the absence of slippage) to rotate clockwise, which will lead to movement of the bicycle from left to right, i.e., against the force $F$. However, the moment from the forces of tension of the chain, which is also applied to the sprocket with the pedals, must also be taken into account. This moment may be directed anticlockwise and may be sufficiently great to exceed the moment from the force $F$ and force the sprocket to rotate anticlockwise. This may lead to movement of the bicycle from right to left, in the direction of the force $F$. Such a situation comes about when the force $F$ is external and the parameters of the bicycle satisfy the condition $\sigma \in(0,1)$. If, however, the force $F$ is internal, then, as shown above, the given situation can never be realized, i.e., the bicycle will move against the direction of the force $F$.

\section{The general problem of controlling the motion of a bicycle in the vertical plane using an internal or external force $F$ applied to the lower pedal statement of the problem of determining the accelerations}

The force $F$ directed along the horizontal and against the $x$ axis (Fig. 1) is applied to the pedal (i.e., to the body 1) either as in Problem 1 or as in Problem 2. Thus, this force is either internal or external. The bicycle at the initial instant $t=0$ is at rest. It is required to determine the dynamics of its motion at fairly short times $t>0,$ on the assumption that the wheels are in contact in a unilateral way with a rough support plane with a Coulomb friction coefficient $f$.

\textbf{Remark 1.} The use of an internal force $F$ in the way indicated above (as in Problem 1) is equivalent to applying a corresponding control moment to the wheel $2,$ which is characteristic of motorcycles, where such a moment is created by the motor.

\textbf{Remark 2.} It can be assumed that the bicycle has completed uniform uncontrolled motion without wheel slip, and that the cyclist has then decided to begin controlled motion in the way indicated above.

\textbf{Remark 3.} Unilateral contact of the wheels with the support plane makes it possible to consider the well-known trick of motion with one of the wheels separated from the support. The situation where both wheels are separated is not ruled out either, but these are different problems.

\subsection{The principal equations of the kinematics and dynamics of motions of the bicycle without separation of the wheels}
\setcounter{equation}{0}
Let the force $F$ be internal (as in Problem 1). We will write three equations of the dynamics of planar motion of the bicycle, taking into account that the centre of mass does not move along the vertical:
\begin{equation}\label{eq21}
a=S_{2}+S_{3} \quad 0=-N_{2}-N_{3}+g
\end{equation}
\begin{equation}\label{eq22}
I_{1} \varepsilon_{1}+I_{2} \varepsilon_{2}+I_{3} \varepsilon_{3}=N_{2} b_{2}-N_{3} b_{3}-\left(S_{2}+S_{3}\right) h_{0}
\end{equation}

Equations (2.1) are the equations of motion of the centre of mass of the bicycle along the horizontal and vertical, while Eq. (2.2) is the equation of change in the angular momentum of the bicycle relative to its intrinsic centre of mass.

We will describe in greater detail how Eq. (2.2) is obtained. Suppose $v_{k}$ is the velocity of the centre of mass of the body $k(k=1,2,3,4)$ It is clear that $v_{k}=v,$ where $v$ is the velocity of the centre of mass of the bicycle, which is directed along the $x$ axis. Let $\mathrm{Cxy}$ be a system of König coordinates ($C$ is the centre of mass of the bicycle), and let $m_{k}$ and $y_{k}$ be the mass and vertical coordinate of the centre of mass of the body $k$ in the $\mathrm{Cxy}$ system. Then, for time derivatives from the kinetic moments of the bodies relative to their general centre of mass $\mathrm{C}$, we will obtain
$$
\dot{K}_{i}=I_{i} \varepsilon_{i}+m_{i} a y_{i}, \quad i=1,2,3 ; \quad \dot{K}_{4}=m_{4} a y_{4}
$$

Summing up these expressions and using the equality
$$
m_{1} y_{1}+m_{2} y_{2}+m_{3} y_{3}+m_{4} y_{4}=0
$$
we will obtain the left-hand part of Eq. (2.2). The right-hand part is obtained in the standard way. We will write equations for the changes in the kinetic moments of the bodies $1,2,$ and 3 separately in relation to their intrinsic centres of mass, taking into account that forces of tension of the chain $T_{1}$ and $T_{2}$ are applied to the body 1, and the same forces but with opposite signs are applied to the body $2.$ We will obtain

\begin{equation}\label{eq23}
I_{1} \varepsilon_{1}=F L+T r_{1}, \quad I_{2} \varepsilon_{2}=-S_{2} R-T r_{2}, \quad I_{3} \varepsilon_{3}=-S_{3} R
\end{equation}

Then, we will write the kinematic equation that follows from the condition of the absence of stretchability of the drive chain between the sprocket 1 and the sprocket of the wheel 2 :
\begin{equation}\label{eq24}
\varepsilon_{1} r_{1}=\varepsilon_{2} r_{2}
\end{equation}

The use of Eqs (2.3) makes it possible to write Eq. (2.2) in the form
\begin{equation}\label{eq25}
N_{3} b_{3}-N_{2} b_{2}+\left(S_{2}+S_{3}\right) h+T\left(r_{1}-r_{2}\right)+F L=0 ; \quad h=h_{0}-R
\end{equation}

Excluding $T$ and $\varepsilon_{1}$ from the obtained equations, we will have the following system of five equations:

\begin{equation}\label{eq26}
\begin{array}{l}
a=S_{2}+S_{3}, I_{0} \varepsilon_{2} / R=F \sigma-S_{2}, I_{3} \varepsilon_{3} / R=-S_{3} \\
N_{2}+N_{3}=g, \quad N_{3} b_{3}-N_{2} b_{2}+\left(S_{2}+S_{3}\right) h+F \sigma R k_{2}+S_{2} R k_{1}=0
\end{array}
\end{equation}
where the following notation is introduced:
\begin{equation}\label{eq27}
I_{0}=I_{1} \mu^{2}+I_{2}, \quad k_{1}=I_{1} \mu(\mu-1) / I_{0}, \quad k_{2}=\left(I_{1} \mu+I_{2}\right) / I_{0}
\end{equation}
and parameter $\sigma$ is defined by the last equality in system (1.3). If the force $F$ is external (as in Problem 2), it is not difficult to show that, instead of system (2.6), we obtain a system differing from system (2.6) in the presence of the term $-F$ in the first equation and the term $-F(L+h)$ in the final equation.

In system $(2.6), S_{j}, N_{j}, \varepsilon_{j}(j=2,3),$ and $a$ are unknown quantities to be determined (in both problems, both for an internal and for an external force $F$).

Thus, for correct and unambiguous solution of system (2.6) or the similar system in the case of an external force, it is necessary to have two more equations, which are obtained additionally from physical considerations following from the proposed possible motion of the bicycle. We will give the additional equations for different cases of possible motions of this kind:
neither of the wheels 2 and 3 slips:
$$
a=\varepsilon_{j} R, \quad j=2,3
$$
the wheel 2 (wheel 3) slips, while the wheel 3 (wheel 2) does not slip:
$$
a=\varepsilon_{5-j} R, \quad S_{j}=-f N_{j} \operatorname{sign} a_{j}, \quad a_{j}=a-\varepsilon_{j} R, \quad j=2 \quad(j=3)
$$

both the wheels 2 and 3 slip:
$$
S_{j}=-f N_{j} \operatorname{sign} a_{j}, \quad j=2,3
$$
Before setting out the solutions of the corresponding equations of the dynamics in the four given cases, we will examine the possibility of static equilibrium of the bicycle with the indicated application of an internal or external force $F$ to the pedal and the presence only of frictional forces of rest at the contact points of the wheels. We will show that for a bicycle (more accurately, for the combination of rigid bodies making up the bicycle, with or without a cyclist, depending on the examined problem) such equilibrium is impossible with any non-zero force $F$ if there are no moments of rolling friction on its wheels. In fact, if the force $F$ is internal, the equations of static equilibrium have the form
\begin{equation}\label{eq28}
\begin{array}{l}
S_{2}+S_{3}=0, \quad N_{2}+N_{3}-g=0, \quad N_{2} b_{2}-N_{3} b_{3}-\left(S_{2}+S_{3}\right) h=0 \\
F L+T r_{1}=0, \quad-S_{2} R-T r_{2}=0, \quad-S_{3} R=0
\end{array}
\end{equation}

from which it follows that $S_{2}=S_{3}=0, T=0,$ and $F=0$, i.e., equilibrium is possible only at zero force $F$. If the force $F$ is external, then the term $-F$ is added to the left-hand part of the first equation of system $(2.8),$ and the term $F(L+h)$ to the left-hand part of the third equation. From these equations it follows that $S_{3}=0, S_{2}=F, T=-F L / r_{1},$ and $T=-F R / r_{2}.$ The last two equations are possible only if $L=R r_{1} / r_{2}.$ For a real bicycle, $r_{1}>r_{2}$, and the refore $L>R$, i.e., the pedal rests on the support plane. Thus, static equilibrium is again impossible. Note that a similar conclusion is also correct when force $F$ is inclined to the horizontal.

\textbf{Remark.} In their problem book (Ref. 15, p. 81), Gnädig et al. give an erroneous solution for the case where the force $F$ is external. This error stems from the $a$ priori assumption that the system of bodies making up the bicycle is in equilibrium. On this basis, three equations of equilibrium are examined for the system as a whole as an absolutely rigid body. However, these equations are the consequence of the indicated assumption and cannot be a criterion for retention of static equilibrium of the examined system of bodies. As can be seen, a more detailed analysis taking into account the specific nature of the examined system of rigid bodies leads to the impossibility of static equilibrium of the bicycle at any force $F$, either internal or external.

\section{Formulating the results for the case of an internal force $F$}
\setcounter{equation}{0}
In this section, we will formulate the results that are obtained by applying to the pedal of the bicycle an internal force $F$ (as in Problem 1 from Section 1) for the four cases indicated above. Everywhere below, unless stipulated otherwise, it is assumed that
$$
\mu=1\left(r_{1}=r_{2}\right), \quad b_{2}=b_{3}=b
$$
Furthermore, to simplify the formulae, it has been assumed that $M=1$ (i.e., the total mass of the bicycle with the cyclist is selected as the unit for measurement of mass). We will introduce the following (already dimensionless) notation:
\begin{equation}\label{eq31}
\begin{array}{l}
M_{0}=1+M_{2}+M_{3}, M_{1}=1+M_{2}, \quad M_{2}=\left(I_{1}+I_{2}\right) / R^{2}, \quad M_{3}=I_{3} / R^{2} \\
b_{0}=f\left(h+M_{0} R\right), \quad b_{1}=f h\left(2 M_{1}\right), \quad b_{11}=f\left(h+M_{1} R\right) \\
\xi_{1}=f g \frac{M_{0}}{2\left(1+M_{3}\right)-b_{0} / b}, \quad \xi_{2}=f g \frac{M_{0}}{2 M_{3}+b_{0} / b}, \quad \xi_{3}=f g \frac{M_{1} b}{b_{11}} \\
\xi_{4}=f g \frac{\eta}{2 \eta-f\left(1+2 M_{2}\right) R}, \quad \xi_{5}=g \frac{\eta}{\left(1+2 M_{3}\right) R}, \quad \xi_{6}=g \frac{\eta}{R}, \quad \eta=b-f h \\
a_{21}=f g \frac{2 b+f R}{b_{11}}\left(\frac{b}{2 b+f R}-\frac{M_{3} b s_{1}}{2 M_{3} b+b_{0}}\right), \quad a_{22}=f g \frac{2 b+f R}{2 M_{3} b+b_{0}}\left(\frac{b}{2 b+f R}+\frac{M_{3} b s_{2}}{b_{11}}\right) \\
a_{31}=f \frac{g b+F \sigma R}{2\left(1+M_{3}\right) b-f h}, \quad a_{41}=f \frac{F \sigma R}{\eta}, \quad s_{1}=\frac{F \sigma-\xi_{3}}{\xi_{2}-\xi_{3}}, \quad s_{2}=\frac{F \sigma-\xi_{2}}{\xi_{3}-\xi_{2}}
\end{array}
\end{equation}

\textbf{Assertion 1.} When the bicycle moves without slip of either wheel, the following conditions are fulfilled:
$$
\text { if } b_{0}<b \text {, then } 0<F \sigma<\xi_{1} ; \text { if } b \leq b_{0} \text {, then } 0<F \sigma<\xi_{2} \text{.}
$$
The acceleration $a=a_{11}$ of the centre of mass of the bicycle in this case is given by formula (1.4).

\textbf{Assertion 2.} To realize motion of the bicycle with slip of the front wheel 3 only (wheel 3 slips forwards), the following conditions are fulfilled (in each of the examined cases, the corresponding acceleration $a$ of the centre of mass of the bicycle is given in parentheses):

if $0<b<b_{1},$ then $\xi_{3}<F \sigma<\xi_{2}\left(a=a_{21}>0\right)$

if $b_{1} \leq b<b_{11},$ then $\xi_{2}<F \sigma<\xi_{3}\left(a=a_{22}>0\right)$

if $b_{11} \leq b<b_{0}$, then $\xi_{2}<F \sigma<\xi_{4}\left(a=a_{22}>0\right)$

if $b>b_{0}$, then this motion is impossible.\\

\textbf{Assertion 3.} When the bicycle moves with slip of the rear wheel 2 only (wheel 2 slips backwards), the following conditions are fulfilled:
$$b>b_{0}, \xi_{1}<F \sigma<\xi_{5}\left(a=a_{31}>0\right),$$ 
if $b<b_{0},$ tnen this motion is impossible.

\textbf{Assertion 4.} When the bicycle moves with slip of both wheels (the wheel 2 in this case slips backwards, and the wheel 3 slips forwards), the following conditions are fulfilled:

if $b_{11}<b<b_{0},$ then $\xi_{4}<F \sigma<\xi_{6}\left(a=a_{41}>0\right)$

if $b_{0}<b<+\infty,$ then $\xi_{5}<F \sigma<\xi_{6}\left(a=a_{41}>0\right)$

if $0<b<b_{11},$ then this motion is impossible. \\

Substantiation of these assertions is given in Section 4.

The combination of the assertions 1 to 4 leads to the following assertion.

\textbf{Assertion 5.} If $0<b<b_{1}$ and $\xi_{3}<F \sigma<\xi_{2}$, then motion occurs either without slip of either wheel or with slip of the front wheel (ambiguity). In all remaining cases, motion is unambiguous in accordance with the results of the assertions 1 to $4,$ or loss of the unilateral constraint occurs for the front wheel.

\textbf{Remark 1.} From Assertion 5 it follows that, when certain conditions are fulfilled, the motion is ambiguous (the Painlevé paradox). The realized motion of the bicycle depends on the prehistory of its movement. Such issues are discussed in detail in other papers by the present author. 16,17

\textbf{Remark 2.} The loss (weakening) of the unilateral constraint for the front wheel of the bicycle occurs at a fairly high force $F$. In this case, the normal reaction on the front wheel becomes negative. Further motions of the bicycle with possible separations of the front wheel should be considered in a separate problem.

\textbf{Remark 3.} If $f \rightarrow \infty$, then only cases of the Assertions 1 and 2 are realized, from which, in part, there follow the results of solving the Problem 1 from Section 1 for the motion of the bicycle without wheel slip. However, the ambiguity (i.e., the additional possibility of motion with slip of the front wheel with certain values of the force $F$ is retained. This was pointed out to the author by F.L. Chernous'ko.

\textbf{Remark 4.} It is surprising that motions of the bicycle with slip of the front (driven) wheel are possible. In fact no moments are applied to this wheel. However, it must be borne in mind that at the centre of the front wheel there acts a horiz ontal force of reaction from the bicycle frame. It is this force that may in certain situations cause slip of the front wheel. A problem of this type for one wheel with a horizontal force applied at its centre is given in Meshcherskii's problem book (Ref. 18, problem 35.4). The complete solution of this problem is available. ${ }^{19}$

\section{Substantiation of the results of Section 3}
\subsection{Proof of Assertion 1}
\setcounter{equation}{0}
The solution of system of equations (2.6) with the corresponding additional equations and notation (3.1) leads to the formulae
\begin{equation}\label{eq41}
\begin{array}{l}
a=\frac{F \sigma}{M_{0}}, \varepsilon_{2}=\varepsilon_{3}=\frac{F \sigma}{M_{0} R} \\
S_{2}=\frac{F_{0}\left(1+M_{3}\right)}{M_{0}}, \quad S_{3}=-\frac{F \sigma M_{3}}{M_{0}} \\
N_{2,3}=\frac{g b_{3} \pm F \sigma d}{2 b}, d=R+\frac{h}{M_{0}}
\end{array}
\end{equation}

For the correct realization of the examined motion, it is necessary to ensure that the inequalities of the unilateral constraints $N_{2}>0$, $N_{3}>0$ are observed, and also the Coulomb inequalities $\left|S_{2} / N_{2}\right|<f$ and $\left|S_{3} / N_{3}\right|<f$ for the frictional forces of rest $S_{2}$ and $S_{3}$. The use of these inequalities together with relations (4.1) and notation (3.1) leads to the results of the Assertion $1.$

\subsection{Proof of Assertion 2}
The solution of system of equations (2.6) with the corresponding additional equations and allowance for notation (3.1) leads to the formulae
\begin{equation}\label{eq42}
a=\frac{F \sigma-f N_{3} \delta_{3}}{M_{1}}, S_{2}=\frac{F \sigma+f N_{3} \delta_{3} M_{2}}{M_{1}}, \delta_{3}=\operatorname{sign} a_{3}
\end{equation}

Then, it is not difficult to estimate that
\begin{equation}\label{eq43}
a_{3}=\frac{F_{\sigma}}{M_{1}}-f N_{3} \delta_{3}\left(\frac{1}{M_{1}}+\frac{1}{M_{3}}\right)
\end{equation}
from which it follows that the value $l_{3}=-1$ cannot be real ized, as in this case, under the conditions $N_{3}>0$ and $F>0,$ we obviously obtain $a_{3}>0,$ which leads to a contradiction when $\mathrm{I}_{3} = sign a_3$.
 
 Thus, only $\mathrm{l}_{3}=+1, \mathrm{a}_{3}>0$ is possible, i.e., the front wheel can only slip forwards over the course of motion along the $x$ axis. Then, we will obtain
\begin{equation}\label{eq44}
N_{2}=\frac{g\left(b-f d_{0}\right)+F \sigma d_{1}}{q}, \quad N_{3}=\frac{g b-F \sigma d_{1}}{q} ; \quad q=2 b-f d_{0}, \quad d_{0}=\frac{h}{M_{1}}, \quad d_{1}=R+d_{0}
\end{equation}

Using the inequality $a_{3}>0$ and the Coulomb inequality for $S_{2}$ and $\mathrm{N}_{2}$, from formulae (4.2) to (4.4) we will obtain
\begin{equation}\label{eq45}
N_{2}>0, \quad N_{3}>0, \quad f N_{3} M_{0} / M_{3}<F \sigma<f N_{2} M_{1}-f N_{3} M_{2}
\end{equation}

$N_{2}$ and $N_{3}$ are defined by formulae (4.4). Then, from relations (4.4) and (4.5) follow the inequalities
\begin{equation}\label{eq46}
f M_{0} \frac{g b-F \sigma d_{1}}{M_{3} q}<F \sigma<f \frac{g(b-f h)+D \sigma d_{1}\left(M_{1}+M_{2}\right)}{q}, 0<\frac{g b-F \sigma d_{1}}{q}<g
\end{equation}

Hence, when $q>0,$ we obtain
\begin{equation}\label{eq47}
F \sigma>f g \xi_{2}, \quad F \sigma\left[2(b-f h)-f R\left(M_{1}+M_{2}\right)\right]<f g(b-f h), \frac{g\left(f d_{0}-b\right)}{d_{1}}<F \sigma<\frac{g b}{d_{1}}
\end{equation}

The magnitude of ${\xi}_{2}$ is given by the second equation in the third line of formulae (3.1). If $\mathrm{q}<0$, then in relations (4.7) all the signs of the inequalities must be changed to the opposite signs. To solve the obtained inequalities (4.7), we will introduce, using notation (3.1), the following functions of parameter $b$:
\begin{equation}\label{eq4.8}
F_{k}(b)=\xi_{k}, \quad k=1,2,3
\end{equation}

In accordance with inequalities (4.7), these functions limit the upper and lower values of EST, depending on the value of the parameter $b \in (0,+\infty)$, and when $b=b_{1}$ they acquire the same value
$$
F_{0}=f g \frac{h}{2\left(M_{1} R+h\right)}
$$

An analysis of inequalities (4.7) with the use of functions (4.8) and their elementary properties leads to the results of the Assertion 2.

\subsection{Proof of Assertion 3}
The solution of system of equations (2.6) with the corresponding additional equations and notation (3.1) leads to the formula
$$
a=-f N_{2} \delta_{2} /\left(1+M_{3}\right), \quad \varepsilon_{2} R=\left(F \sigma+f N_{2} \delta_{2}\right) / M_{2}, \quad\left(\delta_{2}=\operatorname{sign} a_{2}\right)
$$
from which we will obtain the equality
\begin{equation}\label{eq49}
a_{2}=-\frac{F \sigma}{M_{2}}-f N_{2} \delta_{2}\left(\frac{1}{M_{2}}+\frac{1}{1+M_{3}}\right)
\end{equation}

From this, with $\delta_{2}=1$, we arrive at a contradiction, as then, with $N_{2}>0$ and $F>0$, obviously we will have $a_{2}<0.$ Thus, we have $\delta_{2}=-1$ and $a_{2}<0$ (the rear wheel 2 may slide only backwards, against the direction of the $x$ axis), and from the relation (4.9) we will obtain the inequality
\begin{equation}\label{410}
F \sigma>f N_{2} \frac{M_{0}}{1+M_{3}}
\end{equation}

Furthermore, the Coulomb inequality for the force $S_{3}$ leads to the relation
\begin{equation}\label{eq411}
N_{2} M_{3}<N_{3}\left(1+M_{3}\right)
\end{equation}

Then, simple calculations lead to the following formulae for normal reactions:
\begin{equation}\label{eq412}
N_{2}=\frac{g b+F \sigma R}{2 b-f \Delta}, \quad N_{3}=\frac{g b-z f \Delta-F \sigma R}{2 b-f \Delta}, \Delta=\frac{h}{1+M_{3}}
\end{equation}

For the realization of the examined motion, it is still necessary to add the inequalities $N_{2}>0$ and $N_{3}>0,$ which, with account taken of formulae (4.12), leads to the double inequality
\begin{equation}\label{eq413}
0<\frac{g b+F \sigma R}{2 b-f \Delta}<g
\end{equation}

Thus, for the realization of the examined motion, it is necessary to observe the inequalities $(4.10),(4.11),$ and (4.13) with account taken of the formulae (4.12). As a result of simple calculations, we obtain the inequalities of the assertion $3.$

\subsection{Proof of Assertion 4}
The solution of the system of equations (2.6) with the corresponding additional equations and notation (3.1) leads to the formulae

\begin{equation}\label{eq414}
\begin{array}{l}
a_{2}=-\frac{F_{0}}{M_{2}}-f N_{2} \delta_{2}\left(1+\frac{1}{M_{2}}\right)-f N_{3} \delta_{3}, \quad a_{3}=-f N_{2} \delta_{2}-f N_{3} \delta_{3}\left(1+\frac{1}{M_{3}}\right) \\
\delta_{2}=\operatorname{sign} a_{2}, \quad \delta_{3}=\operatorname{sign} a_{3}
\end{array}
\end{equation}

Simple analysis of formulae $(4.14).$ with account taken of the inequalities $\mathrm{N}_{2}>0, \mathrm{~N}_{3}>0,$ and $\mathrm{F}>0,$ leads to the conclusion that only the set $\delta_{2}=-1, \delta_{3}=+1,$ with which the inequalities $a_{2}<0$ and $a_{3}>0$ should be observed, is consistent. Thus, here the rear wheel slides backwards, and the front wheel forwards. Then, from the relations (4.14) we will obtain the inequalities
\begin{equation}\label{eq415}
F \sigma>f N_{2}\left(1+M_{2}\right)-f N_{3} M_{2}, \quad N_{2}>N_{3}\left(1+\frac{1}{M_{3}}\right)>0
\end{equation}

Simple calculations lead to the following expressions for normal reactions:
\begin{equation}\label{eq416}
N_{2}=\frac{g \eta+F \sigma R}{2 \eta}>0, \quad N_{3}=\frac{g \eta-F \sigma R}{2 \eta}>0
\end{equation}

and inequalities (4.15) acquire the form
\begin{equation}\label{eq416}
F \sigma>f \frac{g \eta+F \sigma R}{2 \eta}+f M_{2} \frac{2 F \sigma R}{2 \eta}, \frac{2 F \sigma R}{2 \eta}>\frac{1}{M_{3}} \frac{g \eta-F \sigma R}{2 \eta}
\end{equation}

The parameter $\eta$ is defined by the final equality in the fourth line of formulae (3.1). Considering the two cases $\eta>0$ and $\eta<0,$ from inequalities (4.16) and (4.17) we will obtain the results of Assertion $4.$

\end{document}